\documentclass[usenatbib]{mn2e}
\usepackage{graphicx}
\usepackage{amssymb}
\usepackage{dpfloat}
\usepackage{subfig}
\usepackage{journal_names}

\title[A triaxial Fast Rotator]{The SLUGGS Survey: outer triaxiality of the Fast Rotator elliptical NGC~4473}
\author[C. Foster et al.]{Caroline Foster,$^{1,2}$\thanks{E-mail: cfoster@aao.gov.au} Jacob A. Arnold,$^3$ Duncan A. Forbes,$^4$ Nicola Pastorello,$^4$ 
\newauthor Aaron J. Romanowsky,$^{5,6}$ Lee R. Spitler,$^{2,7}$  Jay Strader$^8$ and Jean P. Brodie$^{6}$
\\
$^1$Australian Astronomical Observatory, PO Box 915, North Ryde, NSW 1670, Australia\\
$^2$European Southern Observatory, Alonso de Cordova 3107, Vitacura, Santiago, Chile\\
$^3$Department of Astronomy and Astrophysics, University of California, Santa Cruz, CA 95064, USA\\
$^4$Centre for Astrophysics \& Supercomputing, Swinburne University, Hawthorn, VIC 3122, Australia\\
$^5$Department of Physics and Astronomy, San Jos\'e State University, One Washington Square, San Jose, CA 95192, USA\\
$^6$University of California Observatories, 1156 High Street, Santa Cruz, CA 95064, USA\\
$^7$Department of Physics and Astronomy, Faculty of Sciences, Macquarie University, Sydney, NSW 2109, Australia\\
$^8$Department of Physics and Astronomy, Michigan State University, East Lansing, Michigan 48824 USA}

\begin{document}

\date{}

\pagerange{\pageref{firstpage}--\pageref{lastpage}} \pubyear{2013}

\maketitle

\label{firstpage}

\begin{abstract}
Systematic surveys of nearby early type galaxies (ETGs) using integral field unit spectrograph (IFU) data have revealed that galaxies can hide interesting structures only visible through kinematic studies. As part of their pioneering work, the ATLAS$^{\rm 3D}$ team have shown that most morphologically elliptical galaxies are \emph{centrally} kinematically disk-like. Hence, while global morphology suggests that ellipticals are ellipsoidal/triaxial in shape, their central kinematics may be consistent with (inclined) oblate systems.

Here we study the Fast Rotator elliptical galaxy: NGC~4473. Using slitlets, we obtain galaxy light kinematics out to unprecedentedly large galactocentric radii (2.5 effective radii). While we confirm the IFU results in the central regions, we find that at large galactocentric radii NGC~4473 exhibits a kinematic transition. In the outskirts, we observe clear minor \emph{and} major axis rotation, a tell-tale sign of triaxiality, which agrees well with the galaxy's Hubble type.

This outer `kinematically distinct halo' (KDH) may be expected from simulations of galaxy formation, and in this system contains around one-third of the stellar light. While this galaxy may be a special case, it suggests further investigation of the outskirts of galaxies is needed to confirm the new paradigm of galaxy classification.
%We conclude that it may be premature to  overturn conventional morphological classifications until more galaxies are mapped out on larger scales.

\end{abstract}


\section{Introduction}

\defcitealias{Emsellem11}{E11}
\defcitealias{Foster11}{F11}
\defcitealias{Proctor09}{P09}

Spiral galaxies are well understood as dominated by thin disks of stars and gas on approximately circular orbits. The intrinsic structure of early type galaxies (ETGs), on the other hand, has proven more elusive, with the standard picture evolving over time in response to new waves of kinematic data. The initial discovery of weak stellar rotation in luminous ellipticals sparked their interpretation as triaxial, pressure-dominated systems that formed through violent relaxation \citep[e.g.][]{Illingworth77,Binney78}. However, stronger rotation was subsequently found at lower luminosities, leading to a recognition of two distinct subclasses of ellipticals (Es, e.g. \citealt{Davies83}; \citealt{Kormendy96}; \citealt{Emsellem07}).

This dichotomy was refined through the exquisite 2D kinematics data of the SAURON spectrograph \citep{Bacon01} and the ATLAS$^{\rm 3D}$ \citep{Cappellari11a} survey, motivating a proposed paradigm shift in the classification of ETGs \citep[][hereafter E11]{Cappellari11b,Emsellem11}. Here the longstanding E and lenticular (S0) morphology types are deemed ``misleading'' and displaced by Slow and Fast Rotators, where the latter are oblate axisymmetric systems encompassing most Es along with S0s, in a continuous family that may even be more closely related to spirals than to Slow Rotator Es \cite[also see][]{Krajnovic13}.

While providing remarkable insights, these conclusions should be tempered by recognition of the limitations of the kinematics data to the central regions of galaxies. The SAURON/ATLAS$^{\rm 3D}$ Fast and Slow Rotator classifications are explicitly made within 0.5 $R_e$ or 1.0 $R_e$ (effective radii). With wider-field spectrographs, examples have been found of dramatic transitions in kinematics not far beyond $1 R_e$ (e.g. \citealt{Coccato09}; \citealt{Proctor09}, hereafter P09; \citealt{Arnold11}; \citealt{Cortesi13}). These findings are in agreement with some theoretical predictions \citep[e.g.][]{Novak06,Hoffman10,Bois11,VeraCiro11} and suggest that ETGs may be made up of two kinematically distinct components. This is reminiscent of the emerging picture of two-phase galaxy formation \citep[e.g.][]{Oser10,Wu12}. For our purposes, we will call these the `central' and `halo' components, corresponding to radii inside and outside of this kinematic transition, respectively. 

Such a view is being explored with ongoing, systematic wide-field surveys, including the SAGES Legacy Unifying Globulars and GalaxieS \footnote{http://sluggs.swin.edu.au/} (SLUGGS, Brodie et al. in prep) survey, which focuses on the chemodynamics of nearby ETGs. One component of SLUGGS is the Spectroscopic Mapping of Early-type Galaxies to their Outer Limits (SMEAGOL), which uses a novel mode of mapping two-dimensional stellar kinematics and abundances to large radii via the Keck/DEIMOS spectrograph.

Here we report interesting results from SLUGGS/SMEAGOL on NGC~4473.  This `prototypical' E5 galaxy has a photometric major axis position angle $PA_{\rm phot}=100$ degrees \citep{deVaucouleurs91,Temi05} and resides in the Virgo cluster at a distance of 15.2~Mpc \citep{Blakeslee09}. It has a circularized $R_e$ of 35.6~arcsec (2.7 kpc) and a luminosity of $M_V = -20.9$ \citep{Kormendy09}. Its \emph{central} kinematics are consistent with that of a Fast Rotator \citepalias{Emsellem11}, although it is a Non Regular Rotator that exhibits two peaks in its velocity dispersion, making it one of the few such galaxies in the ATLAS$^{\rm 3D}$ \citep{Krajnovic11}. \citet{Cappellari07} interpret these ``$\rm2\sigma$''
 peaks as indicating the presence of a counter-rotating stellar disk (30 per cent of the stellar mass). \citet{Bois11} suggested that such $\rm2\sigma$ galaxies may be formed in binary disk major mergers.

This paper focuses on the kinematics of NGC~4473 at large radii ($R\sim2.5R_e$) and is organised as follows: Section \ref{sec:data} presents our data, while our analysis of the stellar light kinematics and its results are presented in Section \ref{sec:analysis}. Discussion and conclusions can be found in Section \ref{sec:conclusions}.

\section{Data}\label{sec:data}

Data acquisition and reduction  follow closely the standard SLUGGS procedure \citep[see e.g.][]{Proctor09,Foster09,Foster11}. Full details will be given in a companion paper (Arnold et al. in prep). Below we give a summary.

%\subsection{Imaging}

%The $gri$ imaging of NGC 4473 was taken with Suprime-cam \citep{Miyazaki02} on Subaru on the night of 2010 April 12. The data are reduced using a modified version of the SDFRED pipeline \citep{Ouchi04}. The final seeing on the reduced $gri$ images is 0.7, 0.6 and 0.7 arcsec, respectively. These are supplemented with existing HST/ACS photometric catalogs \citep[see][]{Spitler06,Strader06}.

%For GC identification, we first model (using {\tt IRAF/Ellipse}) and subtract the galaxy light profiles of NGC~4473 and other nearby large galaxies (i.e. NGC~4477 and NGC~4479) on the Suprime-Cam images.  We allow {\tt Ellipse} to vary the center, position angle and ellipticity.

%At the distance of NGC~4473, GCs are unresolved on the images. Hence, for each image we use {\tt IRAF/DAOPHOT/FIND} to identify point sources deviating by $4\sigma$ from a global background level. Aperture corrections are applied and extended sources removed following \citet{Spitler08b}.  Zeropoints are calculated by bootstrapping to the Sloan Digital Sky Survey DR7 \citep{Abazajian09} photometric system using common unsaturated point sources. Finally, the photometry is Galactic extinction-corrected using \citet{Schlegel98}.

%Our GC candidates identification and selection for spectroscopic follow-up will be presented in Foster et al. (in prep).

%\subsection{Spectroscopy}

Data were obtained on 3 separate nights (2011 March 30, 2012 February 19 and 2012 February 20) using the DEep Imaging Multi-object Spectrograph (DEIMOS) on Keck. A total of 4 individual slit-masks were observed under seeing FWHM between 0.7 and 0.95 arcsec with the usual SLUGGS setup (1200 l mm$^{-1}$ grating centered on 7800 \AA\space and 1 arcsec slit width). Slits were positioned in a semi-random way to principally target globular clusters in the field. A minimum of 4 exposures of 0.5 hours were taken for a total exposure time of 2 hours per mask. One mask was exceptionally observed for an additional 0.75 hours yielding a total of 2.75 hours.

We use the {\sc idl spec2d} data reduction pipeline provided online to reduce the DEIMOS data. 
%The pipeline performs flat-fielding with internal flats, wavelength calibration with ArKrNeXe arcs, as well as local sky subtraction. Standard output includes GC spectra with corresponding variance arrays and subtracted background or `sky' spectra for each slit. 
We follow the Stellar Kinematics using Multiple Slits (SKiMS) technique, described in \citet{Norris08} and \citetalias{Proctor09}, to perform non-local sky subtraction for extracting galaxy light spectra from the subtracted ``background'' spectra. This yields 87 SKiMS spectra of the galaxy light out to $\sim2.5R_e$ with sufficient quality for a kinematic analysis.

%Recession velocities of the GCs are obtained using standard SLUGGS procedure \citep{Pota13}. 
%The {\tt IRAF/RV/FXCOR} routine is used to cross-correlate each spectrum with 13 stellar templates observed with similar instrumental setup. Reliable errors are computed by adding in quadrature the average formal {\tt FXCOR} errors with the standard deviation between templates. 
%Objects with reliable recession velocities are considered to be associated with NGC~4473 if their recession velocity is between 1900 and 2600 km s$^{-1}$, consistent with its systemic recession velocity (2244 km s$^{-1}$). We note that there is a clear gap in the distribution of recession velocities both below and above these limits such that NGC~4473 GCs are very well delineated from Virgo interlopers, Galactic stars and background galaxies. Our final sample contains 110 reliable GC velocities going out to 12 $R_e$.

The stellar light kinematics are obtained using the {\sc pPXF} routine \citep{Cappellari04}. We measure the first four velocity moments (i.e. recession velocity, velocity dispersion, $h_3$ and $h_4$) for all SKiMS spectra. %{\sc pPXF} determines the best kinematic parameters and weighted combination of user-supplied templates minimizing residuals between the science spectrum and the fit. Each fit is then carefully inspected for quality control. 
Uncertainties on the measured velocity moments are propagated using the variance at each pixel. 

\section{Analysis}\label{sec:analysis}

\begin{figure*}
\begin{center}
\includegraphics[width=130mm]{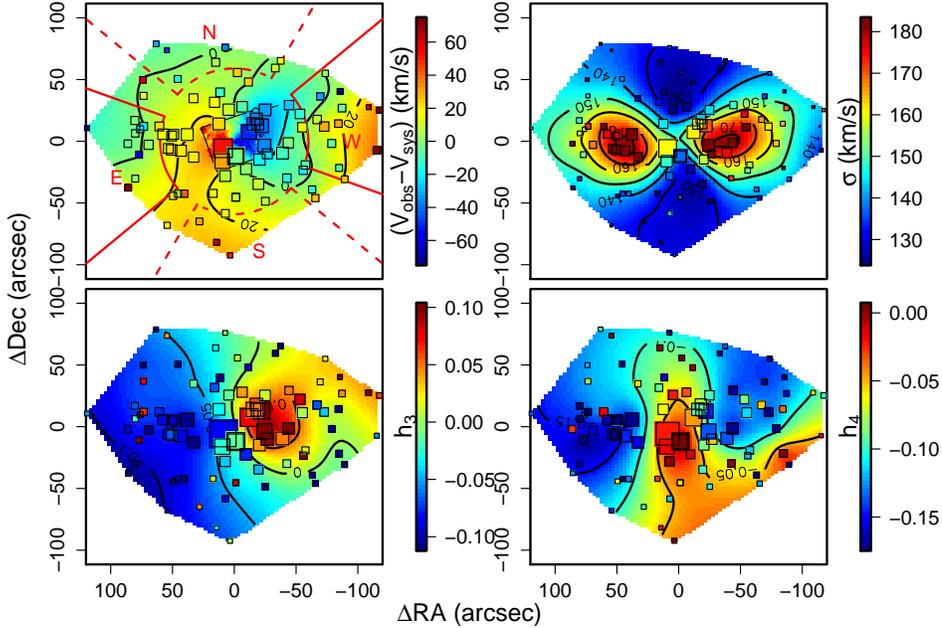}
\caption{Kriging maps of the first four velocity moments of NGC~4473. The relative recession velocity ($V_{\rm obs}-V_{\rm sys}$, top left), velocity dispersion ($\sigma$, top right), skewness ($h_3$, lower left) and kurtosis ($h_4$, lower right) based on kinematics measured on SKiMS spectra from SMEAGOL.  Individual data points are shown as squares with sizes inversely related to the uncertainties. For the upper left panel, the central SAURON \citep{Emsellem04} data are shown within the rectangle and our fiducial major and minor axis `pie' selection regions are shown as red solid and dashed lines, respectively (see text), with red labels. Labelled isovelocity moment contours are shown as solid black lines. North is up, East is left, $R_e=35.6$ arcsec and the photometric position angle is 100 degrees. Several kinematic components are visible and the $2\rm\sigma$ nature of this galaxy is apparent in the velocity dispersion map.}\label{fig:kinmap}
\end{center}
\end{figure*}

%\subsection{Stellar light kinematics}
In order to visualise the results of our sparse SKiMS data on NGC~4473, we use the well established geo-statistical Kriging technique \citep[e.g.][]{Matheron63,Cressie90,Furrer02}. Kriging uses a simple function (polynomial in our case) to fit randomly spaced data of limited sampling without further assumption about the underlying distribution (e.g. symmetry / asymmetry). As such, it is not a physically motivated fit. The technique requires that differences between individual data points are proportional to their mutual distance and that there is indeed a `smooth' and `stable' field to extrapolate.  Our tests show that so long as the underlying field is sufficiently well sampled in regions where steep changes are present, the Kriging technique does a remarkable job of estimating the underlying field (see Pastorello et al., in prep). In practice, we use the {\sc Krig} routine under the {\sc R} package {\sc fields} written by geostatisticians \citep{Furrer02}. The resulting Kriging velocity map based on the SKiMS data of NGC~4473 is shown in Fig. \ref{fig:kinmap}. We wish to emphasize here that no model of the kinematics is used in order to produce these maps, and they are purely uncertainty-weighted 2D fits of the values.

There are several salient features visible in Fig. \ref{fig:kinmap}. First of all, the SKiMS recession velocity analysis nicely complements the published SAURON analysis of the central region as published in \citet{Emsellem04}, including the kinematic transition observed along the major axis near the edge of the SAURON field-of-view \citep[e.g.][]{Krajnovic11}. However, well beyond the SAURON footprint (at $R\gtrsim50$ arcsec), there appears to also be minor axis rotation. Moreover, the inferred counter-rotating component along the major axis seen by \citet{Cappellari07} dominates at large radii (beyond $\sim60$ arcsec). The peaks of the velocity dispersion are also well defined and clearly visible in the map confirming the 2$\sigma$ nature of this galaxy. The peaks in the velocity dispersion correspond to the transition region between the inner and outer (counter-rotating) major axis disks.

Higher order moments ($h_3$ and $h_4$) are less reliably measured in our modest signal-to-noise data. Despite this, clear bulk modulations are clearly visible. The $h_3$ map appears to be anti-correlated with the central major-axis disky component at all radii. The $h_4$ map shows a general trend with overall negative values decreasing with radius along the major axis, indicating that the line profiles become increasingly flattened the further away one moves from the minor axis. This moment is notoriously difficult to measure, especially in our modest data and hence this trend would need confirming using deeper data.

To get a better idea of whether the apparent combination of minor and major axis rotation in the Kriging map at large radii is robust rather than an artifact of the technique itself, it is essential to assess its significance based on direct measures on the data. To that end, we have performed two independent tests.

\subsection{Pie analysis} \label{sec:pie}

\begin{figure*}
\begin{center}
\includegraphics[width=160mm]{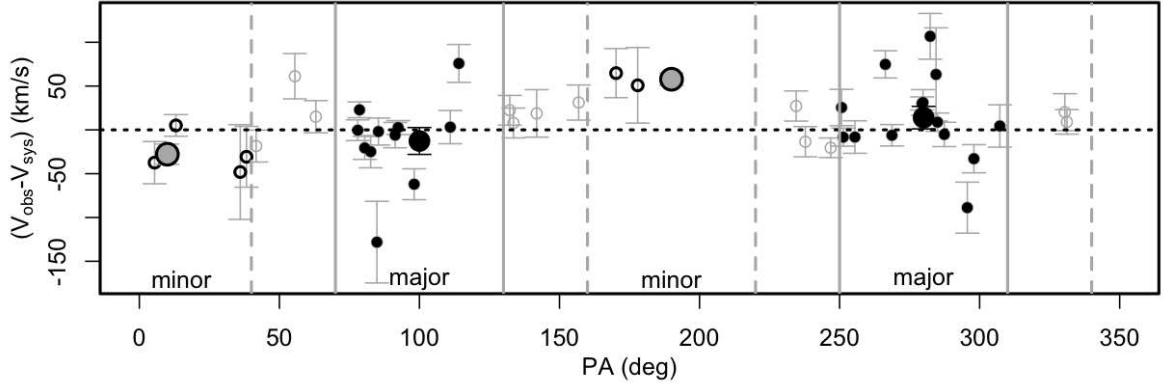}
\caption{Measured relative recession velocity ($V_{\rm obs}-V_{\rm sys}$) as a function of the position angle for data beyond 60 arcmin from the centre of NGC~4473. Our major/minor axis selection is shown as filled/hollow small symbols with means shown as large black/grey symbols with standard error bars. Boundaries of our position angle selections for the major/minor axis are shown as grey filled/dashed lines.}\label{fig:parv}
\end{center}
\end{figure*}

We have selected major and minor axis samples at large R in the following way. We first select data beyond $R_{\rm lim}=60$ arcsec and within $\phi=\pm30$ and $\pm$40 degrees of the major and minor axis (i.e. $PA$=100 and 10 degrees), respectively. These are chosen as our `fiducial' cases for optimal sampling along each axis. Our final conclusions do not depend strongly on these assumptions. Our fiducial `pie' boundaries are shown in Fig. \ref{fig:kinmap} and \ref{fig:parv}. In each pie, we compute the mean normalized recession velocity $\mu_{\rm X}$, where X may be one of N, S, E or W for the Northern, Southern, Eastern and Western pie, respectively. We then compare this mean to that of the opposite pie for both the major and minor axis data (i.e. $\delta_{\rm maj}=|\mu_{\rm E}-\mu_{\rm W}|$ and $\delta_{\rm maj}=|\mu_{\rm N}-\mu_{\rm S}|$, respectively). We assume that rotation along the major or minor axis is the most likely explanation for a non-zero $\delta_{\rm maj}$ or $\delta_{\rm min}$, respectively. We compute the significance of this difference using both Bootstrapping and Monte Carlo methods.

The main advantage of Bootstrapping is that it does not make any assumption about the size and shape of the uncertainties and/or parent distribution. The drawback is that it is only reliable for relatively large samples ($N\gtrsim8)$. We sample with replacement from the major/minor axis sub-samples 50,000 times. Each time, we randomly assign points to either the E or W / N or S pie before computing $\mu_{\rm X}$. In this manner, we obtain an estimate of the distributions of possible $\delta_{\rm maj}$ and $\delta_{\rm min}$ allowed by our data. The resulting p-value ($P_{\rm B}$) quoted in Table \ref{table:stats} is the fraction of values with respective $\delta$ beyond or equal to that measured on the actual data. In other words, it is the probability of randomly measuring such a $\delta_{\rm maj}$ or $\delta_{\rm min}$ in the absence of major or minor axis rotation. As can be seen in Table \ref{table:stats}, $P_{\rm B, maj}=0.044$ and $P_{\rm B, min}=0.009$ in their respective fiducial cases. This means that major/minor axis rotation are detected with 95.6/99.1 percent confidence.

Owing to the possible small sample statistics issues described above, we also use Monte Carlo methods to generate 10,000 random samples of equal size to the measured sample but with values consistent with $(V_{\rm obs}-V_{\rm sys})=0$ km s$^{-1}$ within measurement uncertainties. In order to model these uncertainties for each random datum, we select from a bootstrapped distribution based on sampling all measured uncertainties beyond $R_{\rm lim}$ in our actual data (i.e., without selecting on $\phi$). For each iteration, we randomly assign major/minor axis model data to either the E or W / N or S before computing the $\mu_{\rm X}$. Once again, $P_{\rm MC}$ (see Table \ref{table:stats}) is the fraction of values with respective $\delta$ greater than or equal to that measured on the actual data. In this case, $P_{\rm MC}<0.001$ for both the major and minor axis in their respective fiducial cases. This means that major/minor axis rotation are highly probable at the $>99.9$ percent confidence.

\begin{table}
\begin{tabular}{c c c c c c c}
\hline\hline
$\phi$ & $P_{\rm B, min}$ & $P_{\rm MC, min}$&$N_{\rm min}$& $P_{\rm B, maj}$ & $P_{\rm MC, maj}$&$N_{\rm maj}$\\
(deg)&&&&&&\\
 (1) & (2) & (3) & (4) & (5)&(6)&(7)\\
\hline
40&0.009&0.001&10& 0.162& 0.012& 29\\
30&0.112&0.008&6&0.044&$<0.001$&25\\
\hline
\end{tabular}
\caption{Pie analysis results. These are the Bootstrapping (column 2 and 5) and Monte Carlo (column 3 and 6) significance of the minor and major axis rotation beyond 60 arcsec for two different angular limits (column 1). The total number of data points selected is given in columns 4 and 7 for the minor and major axis rotation, respectively.}
\label{table:stats}
\end{table}

The result of this simple analysis already shows that both major and minor axis rotation are probable at the $>95.6$ and $>99.1$ percent confidence, respectively, depending on the test for our fiducial case.

\subsection{Radial kinemetry}

Our next approach is based on fitting a kinemetric model to the SKiMS data. Kinemetry, analogously to photometry, fits a model to a 2D kinematic maps (see e.g. \citealt{Krajnovic06}; \citetalias{Proctor09}; \citealt{Foster11}, hereafter F11). We closely follow the approach described in \citetalias{Proctor09} and \citetalias{Foster11}. Data are binned using elliptical shells of fixed axis ratio equal to the photometric axis ratio $q_{\rm phot}=0.54$ measured in 2MASS. We select a rolling bin size of 20 data points and fit the simple sine model of a rotating isotropic ellipsoid (or inclined disk) in each bin as described in equations 1 and 2 of \citetalias{Proctor09} with the assumption that the kinematic axis ratio $q=q_{\rm phot}$. We have also tried this with $q=1$ and the results are essentially unchanged. No assumptions are made on $PA_{\rm kin}$, which is allowed to vary freely. Generous limits on $V_{\rm rot}$ are selected such that it may vary freely between $0\le V_{\rm rot}\le150$ km s$^{-1}$. Using chi-square minimisation, we obtain the best-fit uncertainty weighted kinematic position angle ($PA_{\rm kin}$) and rotation amplitude ($V_{\rm rot}$) describing our data in each radial bin. Given that the velocity field of NGC~4473 displays multiple significant components at the same radius, this approach is perhaps too simple.

\begin{figure}
\begin{center}
\includegraphics[width=84mm]{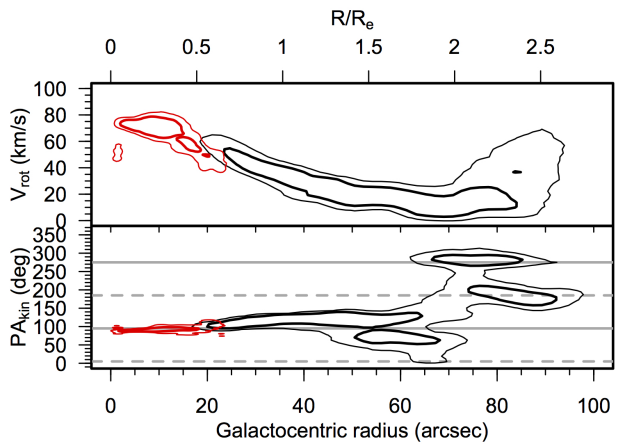}
\caption{The 68 (thick lines) and 95 (thin lines) percent confidence intervals of the circularized radial distribution of $V_{\rm rot}$ (top panel) and $PA_{\rm kin}$ (lower panel) based the kinemetry fits to the SKiMS (black) and SAURON (red) data of NGC~4473. The major/minor axes are shown as horizontal solid/dashed grey lines in the lower panel. Between $70\lesssim R_{\rm circ}\lesssim90$ arcsec both major and minor axis rotation are present.}\label{fig:kinemetry}
\end{center}
\end{figure}

In order to measure the range of possible solutions allowed by the data, we obtain 500 bootstrapped samples based on our SKiMS data by sampling with replacement. We then run our rolling kinemetry algorithm on each of those 500 iterations, obtaining a well-sampled `cloud' of acceptable solutions at each radius. These solutions are presented in Fig. \ref{fig:kinemetry} together with our re-analysis of the SAURON data. We confirm clear central rotation $V_{\rm rot}\gtrsim50$ km s$^{-1}$ along the major axis (i.e. $PA_{\rm kin}\sim110$ degrees). There is a sharp drop in $V_{\rm rot}$ for $40\lesssim R_{\rm circ} \lesssim60$ arcsec beyond which two rotating components are clearly visible. First, a counter-rotating major axis component significant between $60\lesssim R_{\rm circ}\lesssim90$ arcsec with  $V_{\rm rot}\sim 15$ km s$^{-1}$ and $PA_{\rm kin}\sim290$ degrees. The second component corresponds to the minor axis rotation and is significant for $70\lesssim R_{\rm circ}\lesssim95$ arcsec with $V_{\rm rot}\sim 40$ km s$^{-1}$ and $PA_{\rm kin}\sim190$ degrees. These values are in reasonable agreement with the inferred differences in Section \ref{sec:pie}. Hence, there is clear radial overlap of two statistically significant rotating components for the radial range $70\lesssim R_{\rm circ}\lesssim90$ arcsec. This major and minor axis outer rotation does not seem to be simply a reflection of a kinematic twist, but instead the kinemetry solutions flip back and forth between major and minor axis solutions, which suggests two distinct components. We cannot exclude that one of these components may be related to the central major axis rotating component via a large kinematic twist.

Despite this clear kinematic transition, the radially cumulative, luminosity weighted $\lambda_R$ parameter defined by the SAURON team \citep{Emsellem07} is radially constant and very close to its central value of $\lambda_R=0.2$ for $R_{\rm circ}\le110$ arcsec or $R\le2.8R_e$. The $\lambda_R$ parameter is a proxy for the angular momentum per unit mass \citep{Jesseit09}. Adopting the universal limit between Slow and Fast Rotators of $\lambda_{R,\rm lim}=0.1$, NGC~4473 is within the Fast Rotator range even at large galactocentric radii. Using the scale and ellipticity dependent limit ($\lambda_{R,\rm lim}=0.31\sqrt{\epsilon}$, for observed ellipticity $\epsilon_e$) defined in \citetalias{Emsellem11} for $1R_e$, NGC~4473 sits just at the boundary between Fast and Slow Rotators. This suggests that NGC~4473 may be a special case of a 2$\sigma$ Fast Rotator and perhaps not representative of Fast Rotators as a whole.

\section{Discussion and conclusions}\label{sec:conclusions}

%\citetalias{Emsellem11} suggested, while \citetalias{Proctor09}, \citet{Coccato09} and \citet{Cortesi13} have shown that this classification into Slow and Fast Rotators depends on the scale at which the kinematics are probed and hence a galaxy classified based on its central kinematics may not be Slow or Fast Rotators in the outskirts. 

As part of SLUGGS, we have obtained spatially resolved kinematics of the Fast Rotator E galaxy NGC~4473. This galaxy was already well known for its multi-component central kinematics \citep[e.g.][]{Emsellem04, Cappellari05,Cappellari07,Krajnovic11,Krajnovic12}, also shown to be associated with a central light excess \citep{Kormendy09,Dullo13}. In Section \ref{sec:analysis}, we have shown that NGC~4473 exhibits multiple kinematic components with a clear transition occurring at $\sim1.8R_e$. Beyond the transition ($70\lesssim R\lesssim90$ arcsec), two rotating stellar components are visible, the first rotating along the major axis and another rotating along the minor axis. Simultaneous major and minor axis rotation is an unstable configuration in an axisymmetric system and is hence a tell-tale sign of triaxiality \citep[e.g.][]{Franx91,Hoffman09}.

Kinematic transitions are also predicted theoretically. First, disk-disk merger models show that an important kinematic transition is expected beyond 1 $R_e$ \citep[typically beyond the region probed by current IFUs, ][]{Hoffman10}. Also, \citet{VeraCiro11} find that Milky Way type dark matter haloes in the Aquarius Simulations tend to change shape with radius, being more prolate near the center and triaxial further out. Similarly, \citet{Novak06} find that simulated major merger remnants are mostly oblate in their star-dominated centers, and mostly prolate or triaxial in their dark matter dominated outer regions. Using a series of disk-disk mergers, \citet{Bois11} also find that kinematics may change significantly with radius. Hence, we suggest a loose but intuitive decomposition of ETG into two stellar kinematical regions: the central and halo components, corresponding to the regions within and beyond this kinematic transition, respectively. This definition does not exclude that the inner and outer components may be physically  related, but it requires that the inner and outer parts of a galaxy are kinematically significantly different.

The dynamical timescale near the edge of the SKiMS map ($\sim100$ arcsec or 7.6 kpc) can be estimated as $t_{\rm dyn}\sim r/v_c$. We use equation 8 of \citet{Pizzella05} to estimate a circular velocity of $v_c\sim284$ km s$^{-1}$ based on the velocity dispersion ($\sigma$) measured at 100 arcsec (i.e. $\sigma_{100}=150$ km s$^{-1}$). Hence, we estimate a dynamical timescale of $\sim26$ Myr at 100 arcsec for NGC~4473. Given that a typical nearby (redshift $z\sim0$) galaxy on average goes through 1 major merger per ~14 Gyr \citep[e.g.][]{Conselice09}, it is statistically likely that we are seeing a settled halo structure.

%Use equation 8 of Pizzella et al. 2005 V_c=1.35*sigma+81.

%This timescale is t_dyn ~ r/v_c, which for N4473 at 100 arcsec would be ~ 7.5 kpc / 225 km/s ~ 30 Myr (where I have assumed v_c / v_rms ~ 1.5 and v_rms ~ 150 km/s at those radii). Substructures should generally be mixed within a few t_dyn, so let's say ~ 0.1 Gyr. We think it is likely on statistical grounds that the last significant accretion event happened much longer ago than that, which would mean that the halo kinematics represent a "settled" structure, but we currently have no way to be sure.

Therefore, we argue that NGC~4473 has a triaxial stellar `kinematically distinct halo' (KDH) beyond $R\sim 1.8 R_e$. Integrating up the light under the profile given in \citet{Kormendy09} beyond $\sim1.8R_e$, we estimate that this KDH comprises roughly 35 percent of the total stellar light in NGC~4473.

%The outer kinematics are in agreement with what one might naively assume based on its elliptical morphological classification rather than its central kinematics (i.e., NGC~4473 is truly a flattened ellipsoid rather than an inclined disk).

Our results show that NGC~4473 is an interesting example of a triaxial system. %Contrary to previous claims, fast rotators in general may not be characterized overall as simply disklike axisymmetric systems. %Whether NGC 4473 is an exception or the rule may be revealed by careful analysis of the outer regions of a larger sample of galaxies. Such a sample is currently being assembled as part of the SLUGGS and should help verify whether the two-component structure adopted in this work is truly fundamental.
How fundamental the two component structure is, and whether NGC 4473 is an exception or the rule, may be revealed by careful analysis of the outer regions of a larger sample of galaxies.  Such a sample is currently being assembled as part of the SLUGGS survey (Arnold et al. in prep).

\section*{Acknowledgments}
We would like to thank the anonymous referee for his/her careful review and very useful comments on our manuscript. We also acknowledge G. Hau, O. Gerhard and E. Emsellem for helpful discussions. This work was co-funded under the Marie Curie Actions of the European Commission (FP7-COFUND). DF thanks the ARC for support via DP130100388. Supported by the National Science Foundation grant AST-0909237. Based in part on data collected at Subaru Telescope and obtained from the SMOKA, which is operated by the Astronomy Data Center, National Astronomical Observatory of Japan. The data presented herein were obtained at the W.M. Keck Observatory, which is operated as a scientific partnership among the California Institute of Technology, the University of California and the National Aeronautics and Space Administration. The Observatory was made possible by the generous financial support of the W.M. Keck Foundation. The analysis pipeline used to reduce the DEIMOS data was developed at UC Berkeley with support from NSF grant AST-0071048. We acknowledge the usage of the NASA/IPAC Extragalactic Database (NED), which is operated by the Jet Propulsion Laboratory, California Institute of Technology, under contract with the National Aeronautics and Space Administration.

\bibliographystyle{mn2e}
\bibliography{biblio}

\label{lastpage}

\end{document}